\documentclass[runningheads]{cl2emult}

\usepackage{graphicx}
\usepackage{subeqnar}
\usepackage{cropmark}

\begin{document}

%\ifthenelse{\equal{\Ox}{Ja}}{
%\chapter[Noise-Induced Order in Extended Systems]{Noise-Induced Order in Extended Systems:
%A Tutorial}
%\author{Jos\'e M. Sancho}
%\address{Departament d'Estructura i Constituents de la Mat\`eria,
%Univ. de Barcelona,\\ Av. Diagonal 647, E-08028 Barcelona, Spain}
%%\author*{and}
%\author{Jordi Garc\'{\i}a--Ojalvo}
%\address{Departament de F\'{\i}sica i Enginyeria Nuclear,
%Univ. Polit\`ecnica de Catalunya,
%Colom 11, E-08222 Terrassa, Spain}
%}
%{
%%%%%%%%%%%%%%%%%%%%%%%%%%%%
\title*{Noise-Induced Order in Extended Systems:
\protect\newline A Tutorial\footnote{From ``Stochastic Processes in Physics,
Chemistry, and Biology'', J.~A. Freund and T.~P\"oschel (Eds.), Lecture Notes in Physics
vol. 557 (Springer, Berlin, 2000)}}

\toctitle{Noise-Induced Order in Extended Systems: A Tutorial}

\titlerunning{Noise-Induced Order in Extended Systems: A Tutorial}

\author{
Jos\'e M. Sancho\inst{1}
\and Jordi Garc\'{\i}a--Ojalvo\inst{2}
}

\authorrunning{J.M. Sancho and J. Garc\'{\i}a-Ojalvo}

\institute{
Departament d'Estructura i Constituents de la Mat\`eria,
Univ. de Barcelona,\\ Av. Diagonal 647, E-08028 Barcelona, Spain
\and
Departament de F\'{\i}sica i Enginyeria Nuclear,
Univ. Polit\`ecnica de Catalunya,\\
Colom 11, E-08222 Terrassa, Spain}

\maketitle
%}
\begin{abstract}
External fluctuations have a wide variety of constructive effects on the
dynamical behavior of spatially extended systems, as described by
stochastic partial differential equations. A set of paradigmatic situations
exhibiting such effects are briefly reviewed in this paper, in an attempt
to provide a concise but thorough introduction to this active field of
research, and at the same time an overview of its current status. This work
is dedicated to Lutz Schimansky--Geier
on the occassion of his 50th anniversary. Through the years, Prof.
Schimansky--Geier has made important contributions to the field of
spatiotemporal stochastic dynamics, including
seminal investigations in the early 1990's on noise effects in front
propagation, and studies of noise-induced phase transitions and
noise-sustained structures in excitable media, among others.
\end{abstract}

\section{Introduction}

It is well accepted nowadays that noise can have rather surprising and
counterintuitive effects. There are many physical situations in which noise 
exhibits a {\em constructive}, rather than destructive, role in the behavior
of a nonlinear system. A relevant example of this fact, still under active
research nowadays, is the phenomenon of stochastic resonance, in which the
response of a nonlinear system to an external signal under the presence of
fluctuations can be enhanced by tuning the noise intensity to an optimal
(non-zero) value \cite{SG.Gamma98}. In another direction, studies of
stochastic zero-dimensional systems (i.e., systems with only temporal
dependence), reviewed in
\cite{SG.horsthemke84}, showed that noise is able to induce transitions
in such systems. Spatial degrees of freedom provide further interesting
scenarios where the non-trivial role of external noise arises, such as phase
transitions and critical phenomena \cite{SG.Ma76}, and pattern
formation out of equilibrium \cite{SG.cross}.

The present review aims to be a brief introduction to the
influence of external noise on spatially extended, $d$-dimensional
systems (an extensive monograph on the topic can be found in \cite{SG.NISES}).
As far as we know, seminal works in this direction were already carried out in
the late 1970's by Mikhailov \cite{SG.mikhailov79,SG.mikhailov81}. Owing to
the existence of spatial degrees of freedom, the behavior of extended systems
is described by phases in a thermodynamic sense. Under these conditions, the
presence of certain types of external noise affects, in a non-obvious way,
the behavior of the corresponding deterministic (noiseless) system, or even
that of the associated equilibrium system (which has internal noise terms
obeying a fluctuation-dissipation relation). In particular, we
expect not only quantitative changes with respect the deterministic
results, but also qualitatively different, even new, features
{\em induced} by the presence of external fluctuations. In these situations,
na\"{\i}ve predictions based on a deterministic analysis are very far from
giving reliable results. Several examples can be enumerated, some of
which will be reviewed here:
noise-induced spatial patterns
%({\bf NISP}) 
\cite{SG.ojalvo93,SG.BK94,SG.par96,SG.zaikin},
noise-induced ordering phase transitions
%({\bf NIOT})
(both of second order  
\cite{SG.luque94,SG.broeck94,SG.bro94b,SG.genovese},
and of first order \cite{SG.muller97,SG.zaikin99}),
noise-induced disordering phase transitions
%({\bf NIDT})
\cite{SG.broeck94,SG.genovese,SG.ojalvo96},
noise-induced phase dynamics
%({\bf NIPD})
\cite{SG.lacasta},
noise-induced fronts
%({\bf NIF})
\cite{SG.santos98},
and
noise-sustained structures in excitable media
%({\bf NSWSM}) 
\cite{SG.jung95,SG.kadar98,SG.hempel99}.

Most of the noise-induced phenomena enumerated above act in the direction
of enhancing the order in the system. This surprising fact contrasts with
the noise-induced disorder that one could intuitively expect from statistical
mechanics. In the following pages, we review all these ordering effects.
In each of the different physical situations, the notion of ``order'' is
defined, the mechanisms through which noise produces order are discussed,
and a minimal model displaying the phenomenon is given.

%The examples presented above on noise-induced some king of order, on
%the other hand, are counterintuitive phenomena:
%when fluctuations increase, order is surprisingly
%enhanced. In principle, these ordering effects seem to be related
%to the multiplicative character of the fluctuations, as compared 
%to the disordering role of additive fluctuations. But recently this
%simple interpretation has changed; there is an interplay between additive
%and multiplicative noise terms in such a way that additive external noise 
%can order the system \cite{SG.zaikin}. As a consequence, it appears that for
%some regimes
%additive noise can order the system, whereas multiplicative noise can
%disorder it. 
%
%A review of all these striking phenomena is the
%main aim of this contribution. 
 
\section{Noise-Induced Phase Transitions}

The most basic organizing phenomenon in spatially extended systems is the
transition between two macroscopic phases as a certain control parameter
is varied. Such {\em phase transitions} can be characterized by standard tools
in Statistical Mechanics, such as scaling functions, critical exponents,
and renormalization-group transformations \cite{SG.NISES}.

 From a dynamical perspective, the system can be described in a continuous way
by a coarse-grained field $\phi(\vec x,t)$, representing the local density
of a relevant physical variable (e.g., magnetization in a magnetic system,
or relative concentration in a binary alloy). From this viewpoint, a
disordered phase corresponds to the state $\phi(\vec x,t)=0$ (random
distribution of up- and down-spins in magnetic systems, or homogeneous mixture
in alloys), and an ordered phase is given by a non-zero field. As we will see
in what follows, external noise is able to induce phase transitions from
disorder to order. We consider models obeying a stochastic
differential equation of the general form:
\begin{equation}
\frac{\partial \phi(\vec x,t)}{\partial t} = f(\phi)+
g(\phi)\, \eta(\vec x,t) + \nabla^2 \phi + \xi(\vec x,t) \,,
\label{SG.eq:genmod}
\end{equation}
with $\vec x$ defined in a $d$-dimensional space. The additive and
multiplicative gaussian noises have zero mean and correlations: 
\begin{subeqnarray}
& &\langle\xi(\vec x,t)\xi(\vec x',t')\rangle=2\varepsilon\;
\delta(\vec x-\vec x')\;\delta(t-t')\,,
\\
& &\langle\eta(\vec x,t)\eta(\vec x',t')\rangle=
2\;c(\vec x-\vec x')\;\delta(t-t')\,,
\end{subeqnarray}
where $c(\vec x-\vec x')$ is the spatial correlation function of the
multiplicative noise $\eta(\vec x,t)$ [$c(0)$ is proportional to the noise
intensity]. The white additive noise $\xi(\vec x,t)$ is taken to represent
internal fluctuations.

\subsection{A short-time dynamical instability}
\label{sec:stdi}

We now review the most common mechanism of noise-induced ordering
transitions. To do so, we locally analyse the initial evolution of the
system. Averaging (\ref{SG.eq:genmod}) with respect to the probability
density of the two noise terms, and neglecting the diffusive term of the
equation:
\begin{equation}
\left\langle\partial_t \phi\right\rangle =
\langle f(\phi)\rangle+ \langle g(\phi)\, \eta(\vec x,t)\rangle \,.
\label{SG.eq:avmod1}
\end{equation}
We now interpret the multiplicative noise $\eta(\vec x,t)$ in the
Stratonovich sense \cite{SG.gardiner}. This choice is neither arbitrary
nor interested: we simply aim to describe realistic fluctuations, temporally
correlated but with a very small characteristic time.
Under this interpretation, the average of the noise
term appearing in (\ref{SG.eq:avmod1}) can be computed to be
\begin{equation}
\langle g(\phi)\,\eta(\vec x,t)\rangle=c(0)\langle g(\phi)g'(\phi)\rangle\,,
\label{eq:novav}
\end{equation}
where the prime indicates differentiation with respect to the argument
\cite{SG.NISES}.
Coming back to (\ref{SG.eq:avmod1}) one can say that, at the initial
instants of the evolution, fluctuations around the average value of the
field $\langle\phi\rangle$ can be neglected, so that $\langle h(\phi)\rangle
\approx h(\langle\phi\rangle)$ for any function $h$, and one can write
approximately
\begin{equation}
\partial_t \langle\phi\rangle\approx
f(\langle \phi\rangle)+ c(0) g(\langle\phi\rangle)g'(\langle\phi\rangle)
\equiv f_{\rm eff}(\langle\phi\rangle)\,.
\label{SG.eq:avmod2}
\end{equation}
In the zero-dimensional case (no spatial coupling), this approximation is
strictly valid only at short times. At long times, the system heads
towards the steady state, in which the noise effect is in fact
opposite to that of (\ref{SG.eq:avmod2}) (see \cite{SG.broeck96}). In the
spatially extended case, on the other hand, diffusive coupling between
neighbors is able to ``trap'' the system in the short-time dynamics given
by the effective force $f_{\rm eff}$ in (\ref{SG.eq:avmod2}). Hence, a simple
analysis of the zeroes of this function and their stability reveals, in
a na\"{\i}ve but very efficient way, the transition scenario of the system
\cite{SG.note1}.
We will now give some choices of $f$ and $g$, leading to different
noise-induced phase transitions.

\subsection{Noise-Induced Second-Order Phase Transitions}

Let us first consider the following representative model:
\begin{equation}
\partial_t \phi = -\phi(1+\phi^2)+
\phi\, \eta(\vec x,t) + \nabla^2 \phi + \xi(\vec x,t) \,,
\label{SG.eq:mod1}
\end{equation}
which corresponds, in the absence of multiplicative noise, to the well-known
time-dependent Ginzburg-Landau model, frequently used in studies of critical
dynamics \cite{SG.Ma76}, and whose universality class is that of the
Ising model. For the deterministic parameters chosen, the
system resides in the stable disordered state $\phi(\vec x,t)=0$. In the
presence of the multiplicative noise, and according to the discussion of
the previous paragraphs, the effective force (\ref{SG.eq:avmod2}) is equal
to $f_{\rm eff}(\phi)=[c(0)-1]\phi-\phi^3$, which shows that a non-zero
(ordered) value of the field can appear for large enough noise intensities
(for $c(0)>1$ in this approximate analysis). This noise-induced {\em
ordering} phase transition has been observed both numerically
\cite{SG.luque94} and theoretically, through linear stability \cite{SG.BK94}
and mean field \cite{SG.bro94b} analysis. The change of the spatially
averaged field $\frac{1}{V}\int \phi(\vec x,t)\,d\vec x$
at the bifurcation is continuous, which means that the
transition is of second order. The system also displays a subsequent phase
transition back to disorder, also of second order, for larger noise intensities
\cite{SG.ojalvo96}. In spite of the non-equilibrium character of the two
transitions, a finite-size scaling analysis shows that both of them belong to
the equilibrium universality class of the Ising model
\cite{SG.NISES}. The reason for this can be traced back to the existence
of the additive-noise term. In its absence (for which the disordered
phase becomes an absorbing state), a new universality class appears
\cite{SG.grinstein}.

In spite of the simplicity of model (\ref{SG.eq:mod1}), most studies
performed so far on noise-induced phase transitions have been made in 
a model introduced in 1994 by Van den Broeck et al. \cite{SG.broeck94},
which has an additional fifth-order saturating nonlinearity in the
deterministic force, and an external noise with both additive and
multiplicative contributions:
\begin{equation}
\partial_t \phi = -\phi(1+\phi^2)^2+
(1+\phi^2)\, \eta(\vec x,t) + \nabla^2 \phi \,.
\label{SG.eq:mod2}
\end{equation}
The associated effective force is $f_{\rm eff}(\phi)=[2c(0)-1]\phi+
2[c(0)-1]\phi^3-\phi^5$. This system displays again two consecutive
noise-induced phase transitions, an ordering and a disordering one, both
of which belong to the Ising universality class \cite{SG.broeck96}
(which is not surprising, since the relevant terms of the deterministic
hamiltonian are the same as those of the Ginzburg-Landau model). The observed
phenomenology is basically equivalent to that of model (\ref{SG.eq:mod1}).
In particular, for both models the deterministic potential $U_{\rm det}=
-\int f(\phi)\,d\phi$ is monostable, so that the system is in all cases
disordered in the absence of external noise. Moreover, the stochastic
potential\footnote{The
stochastic potential $U_{\rm st}(\phi)$ is defined from the steady-state
probability density $P_{\rm st}(\phi)$ of the zero-dimensional system by means
of $P_{\rm st}(\phi)\sim\exp(-U_{\rm st})$ \protect\cite{SG.broeck96}.}
that can be defined for the local dynamics,
$U_{\rm st}=-\int [f(\phi)-c(0)g(\phi)g'(\phi)]/[c(0)
g^2(\phi)]\,d\phi$,
is also monostable for all $c(0)$ in the two cases, which indicates that
the corresponding zero-dimensional model does not have a noise-induced
transition towards order (towards non-zero field) for any of the two systems.

In its original form, model (\ref{SG.eq:mod2}) cannot distinguish
between the additive and multiplicative contributions of the external
fluctuations to the noise-induced phenomena. Landa et al. \cite{SG.landa98}
slightly modified the model in order to examine the contribution of the
additive-noise term, and discovered the existence of noise-induced phase
transitions (ordering and disordering) controled by additive noise, provided
a multiplicative-noise noise exists.

Another similarity between the noise-induced phase transitions exhibited
by models (\ref{SG.eq:mod1}) and (\ref{SG.eq:mod2}) is that, in both cases,
the mechanism through which noise destabilizes the disordered phase is
linear, as can be easily seen by examining the effective forces in the
two situations. A review of linear instability mechanisms of noise-induced
phase transitions is given in \cite{SG.marta3}. On the other hand, by taking
into account the discussion of Sect. \ref{sec:stdi}, one can
devise in a straightforward way models for which a noise-induced ordering
phase transition is driven by a {\em nonlinear} mechanism. The system
\begin{equation}
\partial_t \phi = -\phi^3(1+\phi^2)+
\phi^2\,\eta(\vec x,t) + \nabla^2 \phi + \xi(\vec x,t)\,,
\label{SG.eq:mod3}
\end{equation}
which has an effective force $f_{\rm eff}(\phi)=[2c(0)-1]\phi^3-\phi^5$,
is an example of this fact. In this case, the destabilization of the
disordered phase $\phi(\vec x,t)=0$ by noise is dynamically nonlinear.

\subsection{Noise-Induced First-Order Phase Transitions}

In all previous examples, the 0-d short-time effective force $f_{\rm eff}
(\phi)$ displays a supercritical pitchfork bifurcation controlled by the noise
intensity, corresponding to a continuous (second-order) phase transition
when spatial coupling is taken into account. Suitable choices of $f(\phi)$
and $g(\phi)$, on the other hand, lead to discontinuous bifurcations in
$f_{\rm eff}(\phi)$, associated to first-order noise-induced phase
transitions. A first simple example is given by
\begin{equation}
\partial_t \phi= -\phi+\phi^3-\phi^5+
\phi\,\eta(\vec x,t) + \nabla^2 \phi + \xi(\vec x,t)\,.
\label{SG.eq:mod5}
\end{equation}
In this case, the effective force can be seen to be $f_{\rm eff}(\phi)=
[c(0)-1]\phi+\phi^3-\phi^5$, which displays a subcritical pitchfork
bifurcation controlled by noise that corresponds to a first-order phase
transition in the spatially extended case. This transition is characterized
by an abrupt change in the spatially averaged field and by a region of
bistability between ordered and disordered states. Such features were
observed by M\"uller et al. \cite{SG.muller97} in a model with the
deterministic force of (\ref{SG.eq:mod5}) and an external noise with
both additive and multiplicative contributions. The separate role of
additive fluctuations was investigated in a modified version of model
(\ref{SG.eq:mod2}), and a first-order phase transition induced by
additive noise was found \cite{SG.zaikin99}, provided multiplicative
noise is present.

The noise-driven discontinuous phase transition displayed by
(\ref{SG.eq:mod5}) is produced by a destabilization of the linear
coefficient of the 0-d effective force.
Similarly to the case of second-order phase transitions, models can be
defined that exhibit noise-induced first-order phase transitions driven
by a nonlinear mechanism. The simplest example is:
\begin{equation}
\partial_t \phi = -\phi(1+\phi^4)+
\phi^2\,\eta(\vec x,t) + \nabla^2 \phi + \xi(\vec x,t)\,,
\label{SG.eq:mod6}
\end{equation}
for which $f_{\rm eff}(\phi)=-\phi+2c(0)\phi^3-\phi^5$. This function
describes a subcritical pitchfork bifurcation at $c(0)=1$ in 0-d, and
hence a nonlinearly-driven discontinuous phase transition can be expected
in the spatially extended case. 

All examples presented in the last two Sections have been founded on
an analysis of the effective force that governs the short-time behavior of
the zero-dimensional system, which we already argued that would become
trapped by spatial coupling in the extended case. The conclusions obtained
by this approach can be confirmed by other methods, such as mean-field
approximations \cite{SG.marta1} and numerical simulations of the complete
models.

\subsection{Noise-Induced Phase Dynamics}

We have just seen that external noise can induce non-equilibrium phase
transitions, both of first and second order, between two {\em stationary}
phases. Now we aim to analyse the {\em dynamical} aspects related to the
appearance of a non-equilibrium phase in the system.
Let us consider a stationary phase, in equilibrium or not, with well-defined
properties (symmetries). If a control parameter, such as temperature, is
suddenly changed, the system undergoes a dynamical process in order to
reach a new steady state corresponding to the new value of the control
parameter. If the initial and final states belong to the same phase, the
dynamical process is simple and linear relaxation dynamics can be used to
model it. On the other hand, if the final state belongs to a phase different
from the initial one, then a pattern dynamics appears. The pattern is
composed of topological defects which evolve towards the new steady state.
This process is controlled by a small number of parameters, and exhibits
dynamics of different universality classes, which are well characterized by
a dynamical exponent and by the scaling properties of the structure function.
Two universality classes of particular interest are those of phase ordering
and phase separation.

%In these two examples the dynamical process takes place in the presence of
%two coexisting phases and as a consequence one can speak of {\em dynamics
%of first order phase transitions}.

In the following paragraphs, we examine whether external noise can induce
these dynamical processes, and what is the resulting dynamical universality
class. The answer to the first question is simple: since external noise can
induce ordering phase transitions, it is also able to induce the corresponding
dynamics. The second question is much more involved. We can advance that the
dynamical universality class is not changed, because the physical growth
mechanisms are the same.
   
\subsubsection{Noise-induced phase ordering.}
This dynamics arises in ferromagnetic systems, when a sudden decrease in
temperature drives the system away from a homogeneous state of zero
magnetization towards a final state of finite magnetization. During this
process, the system exhibits magnetic domains, some of which grow at the
expense of the rest until the whole system presents a homogeneous finite
magnetization. The characteristic size of these domains grows according to
the Allen-Cahn law, $R(t) \sim t^{1/2}$.

A representative model of this situation is that defined by (\ref{SG.eq:mod1}).
We have shown that external noise can induce an ordering phase transition in
that model. Therefore, beyond the transition point, noise can be considered
to drive an initially disordered phase into an ordered state, through the
formation of domains of the two coexisting (positive and negative) ordered
phases and their subsequent dynamics of competition as in equilibrium models.
Numerical simulations confirm this noise-induced dynamics \cite{SG.marta2}.
Moreover, the process has the characteristics of the Allen-Cahn dynamical
universality class: a clear evidence of the scaling behavior of the
structure function with a power law $ \sim t^{1/2}$ is observed. The fact
that the universality class is not changed can be understood because the
driving mechanism of the ordering process is the same as in the deterministic
case: the local curvature of the interface.
 
\subsubsection{Noise-induced phase separation.} 
This universality class arises in homogeneous alloys of two atomic
species\footnote{Phase
separation in fluids is another universality class not discussed here.} 
initially at high temperature. Following a sudden cooling,
the two components of the alloy start to segregate, producing domains rich in
one of the species, whose characteristic size follows the Lifshitz-Slyozov
law, $R(t)\sim t^{1/3}$. The system evolves towards a final state consisting
on two large domains separated by an interphase.

A dynamical model of phase separation can be constructed from model
(\ref{SG.eq:mod1}) if mass conservation is imposed. The explicit model reads:
\begin{equation}
\partial_t \phi = \nabla^2 \left( \phi +
\phi^3 - \nabla^2 \phi + \phi \,\eta({\bf x},t) \right) + \nabla\cdot \vec\xi
(\vec x,t)\;,
\end{equation}
where $\vec \xi(\vec x,t)$ is now a random {\em vector} field. This system
evolves with the restriction that the spatially averaged field,
$\frac{1}{V}\int \phi(\vec x,t)\,d\vec x=\phi_0$, is conserved. As in the
case of model (\ref{SG.eq:mod1}), the external multiplicative noise
$\eta(\vec x,t)$ induces an ordering phase transition, followed by a
disordering one. In the ordered region, domains of the two non-zero phases
coexist and try to grow. However, the dynamics is different from the case
of phase ordering. Owing to the conservation law, domains have to grow at
the expense of other domains, which may be located far away. The driving
mechanism is thus diffusion controlled by local curvature, as in the
deterministic case, and hence we expect the same dynamical universality
class of Lifshitz-Slyozov. Numerical simulations confirm this fact
\cite{SG.marta2}. A scaling behavior of the structure function following
the power law $ \sim t^{1/3}$ is observed.

\section{Noise-Induced Structures}

Besides phase transitions, a second and very important ordering mechanism
in spatially extended systems is that of {\em structure formation}. The
spontaneous appearance and sustenance of spatiotemporal patterns is common
to many non-equilibrium extended media, including hydrodinamical, optical,
chemical and biological systems \cite{SG.cross}. The system is usually
described by a stochastic partial differential equation (or a set of
them), for which a self-sustained non-homogeneous solution is considered
to be an ordered state. For instance, in the case of a single-field
system, an ordered state is represented by a solution $\phi(\vec x,t)$
which depends explicitly on $\vec x$, and which may or may not depend on
time. We will now review the effects of external noise on several different
pattern-forming systems.

\subsection{Noise-Induced Stationary Patterns}

Let us begin by examining whether noise is able to induce stationary
non-homogeneous (ordered) states in a pattern-forming system. The standard
model displaying such a phenomenology is:
\begin{equation}
\partial_t \phi = -\phi(1+\phi^2)+
\phi\, \eta(\vec x,t) - \left(\nabla^2+k_0^2\right)^2 \phi + \xi(\vec x,t) \,,
\label{SG.eq:she}
\end{equation}
which corresponds to the well-known Swift-Hohenberg model, widely used to
describe the formation of stationary structures in hydrodynamics and nonlinear
optics, among other fields. For the parameter region chosen, the stationary
solution of (\ref{SG.eq:she}) in the absence of multiplicative noise is the
disordered state $\phi(\vec x,t)=0$. When the external fluctuations are
considered, the homogeneous solution becomes unstable, as can be seen in
a simple way from a linear stability analysis of the first statistical
moment of $\phi$. To that end, we linearise (\ref{SG.eq:she}), transform
it to Fourier space and average the resulting equation with respect to
both noise distributions, which leads to:
\begin{equation}
\partial_t \langle\phi\rangle = \left[c(0)-1
-\left(k^2-k_0^2\right)^2\right]\langle\phi\rangle\,.
\label{SG.eq:sheav}
\end{equation}
This analysis shows that the disordered phase $\phi(\vec x,t)=0$ becomes
unstable at $c(0)=1$ for a non-zero wavenumber $k_0$, corresponding to
the appearance of a periodic pattern of wavelength $2\pi/k_0$. A similar
conclusion, at least at first order in noise intensity, can be observed when
the stability of higher-order statistical moments is analysed\footnote{This
fact contrasts with the case of 0-d systems, where statistical
moments of different order have different instability thresholds at all orders
\cite{SG.ojalvo96a}.}.

The model of Van den Broeck et al. \cite{SG.broeck94} with the addition of
a Swift-Hohenberg-like spatial coupling term has also been examined in
search of noise-induced patterns. They have been observed coming from
multiplicative \cite{SG.par96} and additive \cite{SG.zaikin} noise. 

\subsection{Noise-Induced Propagation}

The simplest non-stationary phenomenon in a spatially extended system is the
propagation with constant velocity of a structure through the system. We will
see in what follows that external noise is also able to induce such an ordering
effect.

\subsubsection{Noise-Induced Fronts.}
A front is a spatiotemporal structure linking two different homogeneous
states (kink). Experiments have shown that noise supports front propagation
in a chain of bistable diode resonators \cite{SG.locher}.
We now examine the effect of multiplicative noise on the one-dimensional
propagation of a front in the following field equation:
\begin{equation}
\frac{\partial \phi}{\partial t} = \frac{\partial^2 \phi}{\partial x^2}
- \phi(1 + \phi^2)+\phi\,\eta(\vec x,t)\,,
\label{SG.eqfronts}
\end{equation}
In the absence of noise, the homogeneous phase $\phi(x,t)=0$ is the only
stable state of the system. Neither fronts nor any kind of spatial
structure can be sustained at large times; any initial condition
$\phi(x,0)\neq 0$ decays rapidly to the above-mentioned disordered state. 
In the presence of noise, a systematic contribution to the deterministic
dynamics arises as shown in Sect. \ref{sec:stdi}. One can therefore write
an effective model that has the same behavior as (\ref{SG.eqfronts})
on average:
\begin{equation}
\frac{\partial \phi}{\partial t} = \frac{\partial^2 \phi}{\partial x^2}
- \phi(1 -c(0)+ \phi^2)+\xi_{\rm eff}(\vec x,t)\,,
\label{SG.eqfrontseff}
\end{equation}
where the effective noise $\xi_{\rm eff}(\vec x,t)$ has zero mean. According
to this equation, the homogeneous zero state becomes unstable for $c(0)>1$.
Under this condition, an initial localised perturbation produces a
kink-antikink pattern that propagates until a non-zero state $\overline
{\phi}_{st} \sim \sqrt{c(0) -1}$ invades all the system. The mean velocity
of the propagating front can be computed to be $\overline{v} = 2
\sqrt{c(0)-1}$. These analytical predictions can be confirmed by numerical
simulations of the exact model \cite{SG.santos98}.

\subsubsection{Noise-Sustained Signal Propagation.}
Propagating kink-antikink combinations behave as a train of traveling pulses
that can act as information bits in a communication system. In a chain of
asymmetrical double-well oscillators, these pulses are intrinsically unstable
in the absence of noise. Multiplicative noise is able to sustain propagation
of these pulses, as can be understood in a simple way by analysing the
following model:
\begin{equation}
\label{SG.eq:pulse}
\frac{\partial \phi}{\partial t} = \phi(1-\phi)(\phi-a)+
D\frac{\partial^2\phi}{\partial x^2} +v\frac{\partial \phi}{\partial x}
+\phi\;\eta(x,t)\,.
\end{equation}
This equation contains a diffusive and a convective term in order to model
a uni-directional coupling in this one-dimensional system.
For $1/2<a<1$, the deterministic potential is an asymmetric double well,
with $\phi=0$ more stable than $\phi=1$ (which are the two fixed points of
the system). Therefore, in the absence of noise a kink-antikink pulse shrinks
and decays as it propagates. Noise is able to prevent this decay,
as can be seen by writing an effective model with a zero-mean noise, as
has been made in the previous section.
%\begin{equation}
%\label{SG.eq:pulseeff}
%\frac{\partial \phi}{\partial t} = \phi(1-\phi)(\phi-a)+c(0)\phi+
%D\frac{\partial^2\phi}{\partial x^2} +v\frac{\partial \phi}{\partial x}
%+\xi_{\rm eff}(x,t)\,.
%\end{equation}
It is easy to see that an
optimal value of noise intensity renders the effective potential of this
model symmetric \cite{SG.pulse}, which allows sustained propagation of
pulses through the system.

\subsection{Noise-Supported Structures in Excitable Media}

Due to their special characteristics, excitable systems are especially
sensitive to noise. Perturbations due to noise are able, for instance, to
make the system jump from the rest to the excited state. But besides
nucleating excitation pulses, external noise can also sustain their
propagation in the subexcitable regime (in which such motion would not
be possible under deterministic conditions). This fact has been observed
numerically for the case of
spiral waves \cite{SG.jung95}, in what provided the first example, up
to our knowledge, of spatio-temporal stochastic resonance. Additionally,
chemical subexcitable media have also provided the first experimental
evidence of noise-sustained pulse propagation \cite{SG.kadar98}. Such
constructive effects of noise can be modeled by cellular automata
\cite{SG.hempel99}; the obtained results lead also to predictions of
noise-induced synchronization and global oscillations. Another possible
modeling procedure is by means of continuous FitzHugh-Nagumo models of
excitable media. Analyses in this direction have shown that external
multiplicative noise is able to support spiral turbulent states in
simple activator-inhibitor models \cite{SG.ojalvo99}, and that additive
noise induces synchronization phenomena \cite{SG.neiman99}.

\subsection{Other Constructive Effects of Noise}

There are several other examples of noise-induced order in extended media
that have not been described in the previous pages. We have already briefly
mentioned the phenomenon of spatiotemporal stochastic resonance in excitable
media \cite{SG.jung95}. In that case, a certain non-zero but finite value of
noise intensity exists for which the propagation of excitation waves is
optimal. The extension to spatially extended systems of the phenomenon of
stochastic resonance can be understood in other ways. In its purest
interpretation, it corresponds to the enhanced response of a spatially
bistable system (i.e., whose Lyapunov functional has two minima corresponding
to two stable spatiotemporal states) to a harmonic signal \cite{SG.wio}.

Another constructive effect of noise, in this case not as counterintuitive,
is the sustenance of drifting structures is systems with a convection term
\cite{SG.deissler,SG.palma2}. In this case, structures would escape through
the system boundaries, swept by convection, in the absence of noise.
Fluctuations ensure a continuous creation of structures, which are therefore
sustained by noise.

Noise has also been seen to have a constructive influence in globally
coupled systems. As an example, a model of interacting Brownian particles
in a periodic potential has been recently found to exhibit a noise-induced
phase transition \cite{SG.reimann99}, in which the ordered phase can be
interpreted as a ratchet-like transport mechanism, even though the underlying
potential is symmetric.

\section{Conclusions}

We have tried to review, in a clear and pedagogical way, the ordering
influence that external noise exerts in the spatiotemporal dynamics of
extended media. Two main topics have been considered, namely noise-induced
phase transitions and noise-induced structure formation. In each situation,
a specific notion of order has been introduced. In the case of phase
transitions, order is understood in a coarse-graining sense, so that it
corresponds to a non-zero (even uniform) field. In the case of
pattern formation, order is defined in opposition to uniformity, so
that it corresponds to a non-uniform field profile, either depending or
not on time. In any of the two cases, external noise can be seen to induce
order. In each particular situation, the ordered phase can be characterised
by standard tools used in the corresponding deterministic (or equilibrium)
phenomenology. We interpret multiplicative noise in the Stratonovich sense,
impelled by a search of realistic modeling of the fluctuations. Phase
transitions induced by multiplicative noise in the Ito interpretation
have been recently found \cite{SG.munoz}, but they have a disordering
character. Finally, we should also note that all the phenomena reviewed
here can be explained by the short-time dynamical instability mechanism
described in Sect. \ref{sec:stdi}. But ordering transitions exist that
are driven by a different mechanism, an investigation of which is currently
in progress \cite{SG.marta4}.

\section*{Acknowledgments}

We acknowledge financial support from the Direcci\'on General de Ense\~nanza
Superior (Spain), under projects PB96-0241 and PB98-0935. J.G.O. is pleased
to thank Prof. Lutz Schimansky-Geier, to whom this work is dedicated, for
fruitful collaboration in recent years on this topic.


\begin{thebibliography}{99}
\bibitem{SG.note1} It should be noted that this short-time
instability approach does not account for the quantitative influence
of the spatial coupling strength, since this quantity does not appear
explicitly in the analysis given in
(\protect\ref{SG.eq:avmod1})-(\protect\ref{SG.eq:avmod2}).
To take such an effect into account, a mean-field approach can be used
\protect\cite{SG.bro94b,SG.marta1}.
\end{thebibliography}
\end{document}